\documentclass[useAMS,usedcolumn,usegraphicx,usenatbib]{mn2e}

\usepackage{times}
\usepackage{amsmath}                  
\usepackage{amssymb}                  
\usepackage{graphicx}
\usepackage{units}
\usepackage{dcolumn}
\usepackage{longtable}
\usepackage{lscape}
\usepackage{booktabs}
\usepackage{natbib}
\usepackage{aas_macros}
\usepackage[draft,pdftex,pdfpagemode={UseOutlines},bookmarks,bookmarksopen,colorlinks,linkcolor={blue},citecolor={green},urlcolor={red}]{hyperref}

\makeatletter
\newcolumntype{d}[1]{>{\DC@{,}{{.}}{#1}}c<{\DC@end}}
\newcolumntype{o}[1]{>{\DC@{+}{\pm}{#1}}c<{\DC@end}}
\makeatother
\makeatletter
\newcolumntype{f}[1]{>{\DC@{p}{\ldots}{#1}}c<{\DC@end}}
\makeatother

\title[The origin of the Guitar pulsar]{The origin of the Guitar pulsar}
\author[N. Tetzlaff, R. Neuh\"auser and M. M. Hohle]{N. Tetzlaff$^{1}$\thanks{E-mail:
nina@astro.uni-jena.de}, R. Neuh\"auser$^{1}$ and M. M. Hohle$^{1,2}$\\
$^{1}$Astrophysikalisches Institut und Universit\"ats-Sternwarte Jena, Schillerg\"asschen 2-3, 07745 Jena, Germany\\
$^{2}$Max-Planck-Institut f\"ur extraterrestrische Physik, Giessenbachstra{\ss}e, 85741 Garching, Germany}

\begin{document}

\date{Accepted 2009 October 05. Received 2009 October 05; in original form 2009 September 18}

\pagerange{\pageref{firstpage}--\pageref{lastpage}} \pubyear{2009}

\maketitle

\label{firstpage}

\begin{abstract} 

Among a sample of 140 OB associations and clusters, we want to identify probable parent associations for the Guitar pulsar (PSR B2224+65) which would then also constrain its age. For this purpose, we are using an Euler-Cauchy technique treating the vertical component of the Galactic potential to calculate the trajectories of the pulsar and each association into the past. To include errors we use Monte-Carlo simulations varying the initial parameters within their error intervals. The whole range of possible pulsar radial velocities is taken into account during the simulations.\\
We find that the Guitar pulsar most probably originated from the Cygnus OB3 association $\approx\unit[0{.}8]{Myr}$ ago inferring a current radial velocity of $v_r\approx\unit[-30]{km/s}$, consistent with the inclination of its bow shock.

\end{abstract}



\section{Introduction}

There have been several attempts of identifying parent associations of neutron stars \citep[recently e.g.][thereafter Paper I]{2001A&A...365...49H,2008AstL...34..686B,2009AstL...35..396B,2009A&A...497..423M,MNRAS_M7}. Due to large uncertainties of the observables, mainly parallax or distance, and the unknown radial velocity, such investigations are statistical adventures requiring Monte-Carlo simulations to treat the errors. Unlike previous authors we do not initially define a small interval for the radial velocity but cover a wide spectrum instead.\\
The probably most spectacular neutron star in terms of its motion is PSR B2225+65, the pulsar that creates the so-called Guitar nebula, a bow shock generated by the enormous speed indicated by the large transverse velocity component of $>\unit[1500]{km/s}$ at a distance of $\approx\unit[1{.}9]{kpc}$ from the Sun (\citealt{1993MNRAS.261..113H}; \citealt{1993ApJ...411..674T}; ATNF database\footnote{\href{http://www.atnf.csiro.au/research/pulsar/psrcat/}{http://www.atnf.csiro.au/research/pulsar/psrcat/}}; \citealt{2005AJ....129.1993M}). In contrast to that, the pulsar shows quite ordinary properties of a radio pulsar such as no timing anomalies \citep{2007A&A...467.1209H,2008A&A...490L...3B}. Unlike the ``Magnificent Seven'' (radio-quiet soft X-ray sources, \citealt{2001ASPC..234..225T}) which have been discussed in Paper I, the characteristic spin-down age of the Guitar pulsar of $\unit[1{.}1]{Myr}$ (\citealt{2004MNRAS.353.1311H}; ATNF database) is also consistent with cooling models. Thus, we expect to find its kinematic age close to that value. In this \textit{Letter} we investigate a total number of 140 OB associations and clusters (Paper I) kinematically to find associations which are likely to have hosted the supernova which formed the Guitar pulsar. After selecting the candidate birth places we are able to give constraints on the radial velocity.


\section{Procedure}

Different approaches have been made to calculate past trajectories of objects starting with the traditional method of straight lines through space. The effect of the galactic potential can be included utilising either numerical methods such as a Runge-Kutta algorithm \citep[e.g.][]{1999A&A...350..434A,2001A&A...365...49H} or the classical epicycle approximation \citep{1959...Lindblad,1982lbg6.conf..208W}.\\
The latter holds only if the velocity of an object with respect to the Local Standard of Rest (LSR) is sufficiently low. For that reason, since the Guitar pulsar is the fastest pulsar known, we use a different approach, applying a simple Euler-Cauchy algorithm with a fixed time step of $10^4$ years to include only the vertical component of the galactic potential which was adopted from \citet{2003A&A...404..519P} (and references therein). Utilising this technique is fully sufficient for a treatment of some million years and consistent with results applying more complex methods such as a Runge-Kutta numerical integration method applying a more complicated potential such as given in \citet{2001AJ....122.1397H} and references therein \citep[cf. comparison of methods in][]{2009DiplA...Nina}.\\

To investigate the origin of PSR B2224+65 we used the same procedure as we did in our previous paper (Paper I) which has been suggested by \citet{2001A&A...365...49H}. Applying Monte-Carlo simulations varying the parameters within their error intervals, we calculated the trajectories of the pulsar and any association centre into the past and simultaneously their separation at every time step. We then find the minimum separation and the associated time in the past. Since the radial velocity of neutron stars is not directly measurable so far, we vary the radial velocity randomly between $-1500$ and $\unit[1500]{km/s}$. This is somewhat different to our previous work (Paper I) where we used the distribution for pulsar space velocities from \citet{2005MNRAS.360..974H} to derive a radial velocity distribution. Since the Guitar pulsar already shows a transverse velocity of $\approx\unit[1600]{km/s}$ this approach would not cover a wide spectrum of radial velocities.\\
We take the pulsar position from \citet{2004MNRAS.353.1311H} and its proper motion from \citet{1993MNRAS.261..113H}. \citet{1993ApJ...411..674T} give a current distance of $\unit[2]{kpc}$. We use this distance and the distance implied from dispersion measure (galactic electron density distribution model from \citealt{2002astro.ph..7156C}) of $\unit[1{.}86]{kpc}$ \citep{2005AJ....129.1993M} to determine the
parallax (derived from the mean distance of $\unit[1{.}92]{kpc}$) and its error (note that this error does not include measurement uncertainties but only corresponds to the maximum deviation between the parallax derived from the mean distance and
its lower and upper limit derived from the two cited distance values, those values are given without errors in the literature cited; see also \autoref{foot:dist}):

\begin{equation}
	\begin{array}{l c l}
	\alpha &=& 336^\circ\hspace{-0.8ex}.47,\ \delta\ =\ 65^\circ\hspace{-0.8ex}.59,\\
	\pi &=& \unit[0{.}52\pm0{.}03]{mas},\\
	\mu_{\alpha}^* &=& \unit[144\pm3]{mas/yr},\ \mu_{\delta} \ =\ \unit[112\pm3]{mas/yr},
	\end{array}
\end{equation}

\noindent where $\mu_{\alpha}^*$ is the proper motion in right ascension corrected for declination. Heliocentric coordinates and velocity components of our sample of 140 OB associations and clusters (we use the term ``association'' in the following for both) are available in the Appendix of Paper I.


\section{Results}

We initially perform $100{,}000$ Monte-Carlo runs for each of the 140 associations in our sample to find those for which close encounters with the Guitar pulsar are possible in the past $\unit[5]{Myr}$. Then we select those associations for which the smallest separation between its centre and the pulsar found at some time in the past was within three times the association radius or smaller than $\unit[100]{pc}$ and repeated the procedure carrying out another one million runs. \autoref{tab:Guitar_smallestsep} lists those 9 associations for which the smallest separation found (column 2) during the second cycle of investigation was consistent with the association boundaries.\footnote{Increasing the parallax error even by a factor of ten does not affect the results significantly. Four additional associations with small possible separations between their centres and the pulsar are found (Vul OB4, Cyg OB9, Cyg OB7, Cep OB2); however the current distance of the pulsar needed is always found to be less than $\unit[1{.}4]{kpc}$ which is much smaller than dispersion measure implies ($\approx\unit[2]{kpc}$). Moreover, even if we follow the suggestion of \citet{2004ApJ...600L..51C} that the pulsar could have a distance of only $\unit[1]{kpc}$, the radial velocities needed are found to be $>\unit[500]{km/s}$ which is highly unlikely regarding the large transverse velocity of the pulsar.\label{foot:dist}}

\begin{table}
\centering
\caption{Associations for which the smallest separation between the Guitar pulsar and the association centre found (min. $d_{min}$) after 1 million runs was within the radius of the association. Column 2 gives the smallest separation, column 3 the radius of the association (see Appendix of Paper I for references).}\label{tab:Guitar_smallestsep}
\begin{tabular}{l d{2.2} d{2.2}}
\toprule
Association	&		\multicolumn{1}{c}{min. $d_{min}$ [pc]} & \multicolumn{1}{c}{$R_{Assoc}$ [pc]}\\\midrule
Vul OB1			&		1,4																			& 115\\
NGC 6823		&		0,2																			& 34\\
Cyg OB3			&   0,8																			& 53\\
NGC 6871		&		0,3																			&	7\\
Byurakan 1	&		0,5																			& 6\\
Byurakan 2	&		0,8																			&	6\\
NGC 6883		&		1,7																			&	4\\
Cyg OB1			&		2,1																			&	71\\
Cyg OB8			&		0,8																			&	39\\
\bottomrule
\end{tabular}
\end{table}

\begin{table*}
\centering
\caption{Potential parent associations of the Guitar pulsar (PSR B2224+65).\newline
Columns 2 and 3 mark the boundaries of a 68\% area in the $\tau$-$d_{min}$ contour plot for which the current neutron star parameters (columns 4 to 7, radial velocity $v_r$, proper motion $\mu^*_{\alpha}$ and $\mu_{\delta}$ and parallax $\pi$) were obtained and columns 8 to 10 indicate the distance to the Sun $d_{\odot}$ and equatorial coordinates (J2000.0) of the potential supernova. Underneath, we give present positions (equatorial coordinates, distances $d$) and radii $R$ for the respective associations (see Paper I for the complete data including kinematics).}
\label{tab:detailed_scan_Guitar}
\setlength\extrarowheight{3pt}
\small
\begin{tabular}{c f{2.2} f{4.4} >{$}r<{$} o{4.2} o{4.2} >{$}r<{$} >{$}c<{$} >{$}r<{$} >{$}r<{$}}
\toprule
Association	&		\multicolumn{1}{c}{$d_{min}$}	&	\multicolumn{1}{c}{$\tau$}		&	\multicolumn{1}{c}{$v_r$}			& \multicolumn{1}{c}{$\mu_{\alpha}^*$} & \multicolumn{1}{c}{$\mu_{\delta}$} &	\multicolumn{1}{c}{$\pi$}						&	d_{\odot} 	& \multicolumn{1}{c}{$\alpha$}	& \multicolumn{1}{c}{$\delta$}\\
						&		\multicolumn{1}{c}{[pc]}   		& \multicolumn{1}{c}{[Myr]}			&	\multicolumn{1}{c}{[km/s]}		& \multicolumn{1}{c}{[mas/yr]}			& \multicolumn{1}{c}{[mas/yr]}			& \multicolumn{1}{c}{[mas]}						& \multicolumn{1}{c}{[pc]} & \multicolumn{1}{c}{[$^\circ$]}  & \multicolumn{1}{c}{[$^\circ$]}\\\midrule
Vul OB1$^\mathrm{a}$				&		28p110	&	1{.}06p1{.}18		&		193^{+70}_{-73}	&	144+3	&	112+3	&	0{.}52^{+0{.}02}_{-0{.}02}	&	2477\ldots2630	&	295{.}82^{+1{.}00}_{-0{.}66}	& 23{.}3\ldots26{.}2\\
NGC 6823$^\mathrm{b}$						& 25p107			& 0{.}95p1{.}07				& 349^{+102}_{-75}	& 144+3	& 112+3	& 0{.}52^{+0{.}02}_{-0{.}01}& 2230\ldots2390		&	295{.}97^{+0{.}83}_{-0{.}67}	& 23{.}5\ldots26{.}3\\
Cyg OB3$^\mathrm{c}$ 				& 20p65			& 0{.}74p0{.}82						& -27^{+81}_{-70}	& 144+3	& 112+3	& 0{.}52^{+0{.}01}_{-0{.}01}& 2285\ldots2385	& 302{.}08^{+0{.}84}_{-0{.}53}	& 36{.}7\ldots38{.}2\\
Cyg OB1$^\mathrm{d}$							& 45p82			& 0{.}50p0{.}57		&	867^{+167}_{-143} & 144+3 & 111+3 & 0{.}52^{+0{.}02}_{-0{.}02}	& 1640\ldots1760			& 303{.}35^{+0{.}55}_{-0{.}52}	& 39{.}54^{+0{.}26}_{-0{.}30}\\
\bottomrule
\multicolumn{10}{l}{$^\mathrm{a} \alpha = 295^\circ\hspace{-0.8ex}.8,\ \delta = 23^\circ\hspace{-0.8ex}.3,\ d = \unit[1{.}6]{kpc},\ R = \unit[115]{pc}$ \citep{1989AJ.....98.1598B,2001AstL...27...58D}}\\
\multicolumn{10}{l}{$^\mathrm{b} \alpha = 295^\circ\hspace{-0.8ex}.8,\ \delta = 23^\circ\hspace{-0.8ex}.3,\ d = \unit[2{.}3]{kpc},\ R = \unit[34]{pc}$ \citep{2007AJ....133.1092W}}\\
\multicolumn{10}{l}{$^\mathrm{c} \alpha = 301^\circ\hspace{-0.8ex}.7,\ \delta = 35^\circ\hspace{-0.8ex}.9,\ d = \unit[2{.}3]{kpc},\ R = \unit[53]{pc}$ \citep{1989AJ.....98.1598B,2001AstL...27...58D}}\\
\multicolumn{10}{l}{$^\mathrm{d} \alpha = 304^\circ\hspace{-0.8ex}.7,\ \delta = 38^\circ\hspace{-0.8ex}.0,\ d = \unit[1{.}7]{kpc},\ R = \unit[71]{pc}$ \citep{1989AJ.....98.1598B,2001AstL...27...58D}}
\end{tabular}
\end{table*}

\begin{figure}
\includegraphics[width=0.45\textwidth,viewport= 85 265 480 590]{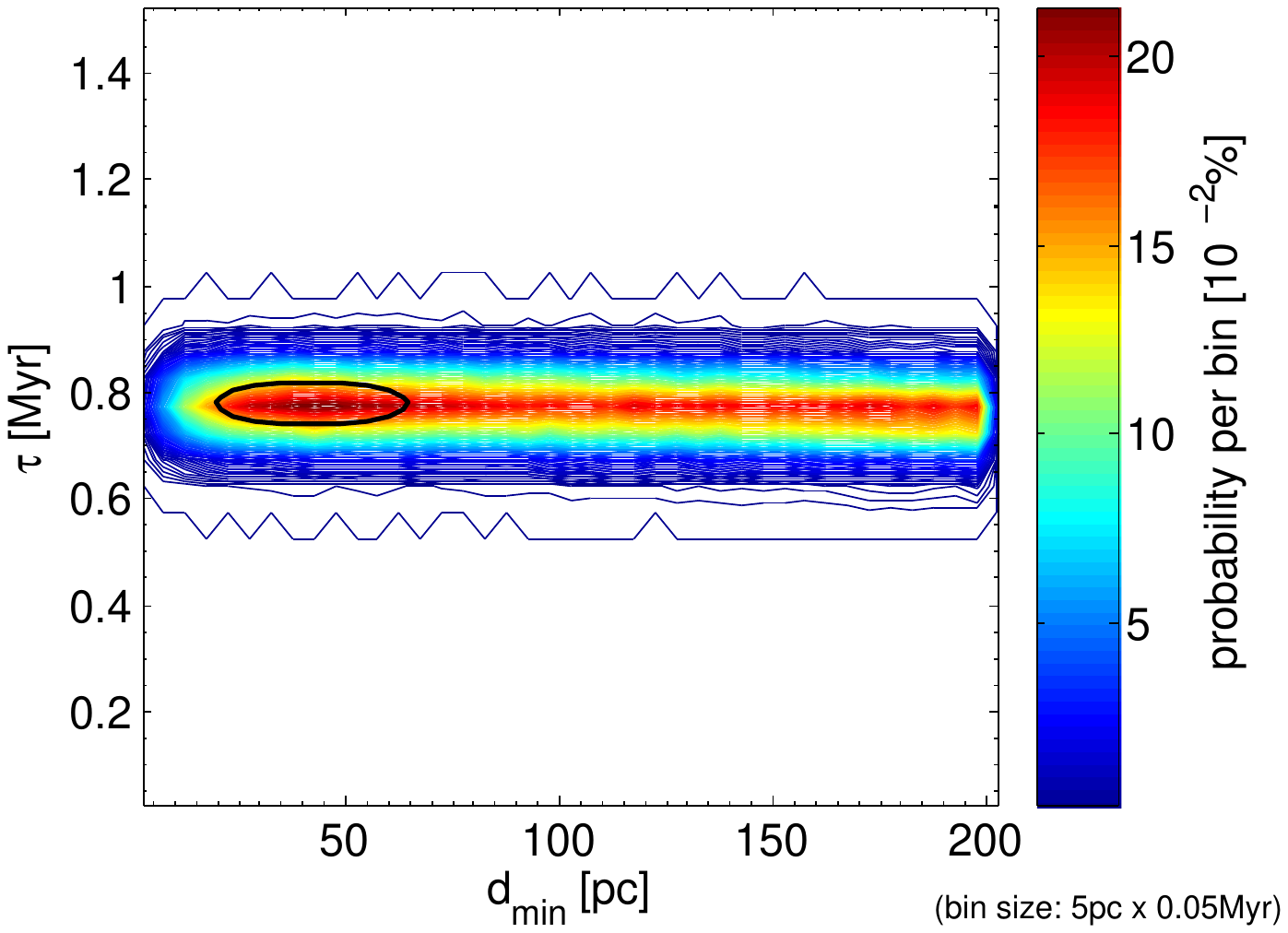}
\caption{Distribution of the number of runs in the $\tau$-$d_{min}$ space for the Guitar pulsar and the centre of Cyg OB3. The ellipse marks the region for determination of the parameters given in \autoref{tab:detailed_scan_Guitar}.}
\label{fig:contour_CygOB3}
\end{figure}

Four of them show a peak in their $\tau$-$d_{min}$ contour plot that is consistent with the respective association radius. \autoref{tab:detailed_scan_Guitar} gives this region of higher probability (columns 2 and 3; boundaries were chosen such that they reflect an approximate 68\% decline from the peak -- see \autoref{fig:contour_CygOB3} as an example) along with present-day parameters which the Guitar pulsar would have if it was born in the particular association. The last three columns indicate the distance of the potential supernova to the Sun as well as its equatorial coordinates. For the definition of selecting the values given in columns 4 to 10 we refer to our Paper I, section 5. It should just be mentioned that the errors given specify 68\% intervals, however are not $1\sigma$ errors.\\

Owing to its large transverse velocity we expect the radial velocity to be very small. Infact, \citet{2004ApJ...600L..51C} investigated the bow shock which the pulsar is generating due to its high speed motion and measured an inclination of $\unit[90\pm20]{degrees}$. This means that the radial velocity is nearly zero. Indeed, we find an association in our \autoref{tab:detailed_scan_Guitar} which is consistent with that: If the pulsar originated from the Cygnus OB3 association (Cyg OB3) its radial velocity would be $\unit[-27^{+81}_{-70}]{km/s}$. For Vulpecula OB1 (Vul OB1) the current radial velocity would also be relatively small ($\approx \unit[190]{km/s}$); however this is significantly larger than for Cyg OB3. The other two parent association candidates infer a substantial larger radial velocity. We thus conclude that Cyg OB3 is most probably the birth association of the Guitar pulsar.\\
Note that NGC 6871, Byurakan 1, Byurakan 2 and NGC 6883 are clusters associated with Cyg OB3. They also appear in \autoref{tab:Guitar_smallestsep}. Their distances to the Sun are still under discussion so that they cover a range between $1{.}6$ to $\unit[2{.}3]{kpc}$. Since NGC 6871 is generally presumed to be the nucleus of Cyg OB3, we repeat our calculations for NGC 6871 adopting the distance of Cyg OB3 ($\unit[2{.}5]{kpc}$). The results do not change significantly: The smallest separation $d_{min}$ found is $\unit[0{.}4]{pc}$ and the peak in the $\tau$-$d_{min}$ contour plot is located outside the boundary of NGC 6871 (radius of $\unit[7]{pc}$). We conclude that the Guitar pulsar was born somewhere within the Cyg OB3 associaton $\approx\unit[0{.}8]{Myr}$ in the past. \\
Given the age of Cyg OB3 of $8$ to $\unit[12]{Myr}$ \citep{1989AN....310..273R,2001A&A...371..675U,2004MNRAS.351.1277S} we can derive the mass of the progenitor star of the Guitar pulsar (the difference between the association age and the kinematic neutron star age yields the progenitor lifetime) of $\unit[15]{M_{\odot}}$ (B1 on the main sequence) (for an association age of $\unit[12]{Myr}$) to $\unit[37]{M_{\odot}}$ (O6 on the main sequence) (for an age of Cyg OB3 of $\unit[8]{Myr}$) under the assumption that all association members formed at once, see \autoref{tab:massage_Guitar}. There is some dispute about the association age since it contains younger clusters (e.g. NGC 6871 is $2$ to $\unit[5]{Myr}$ old according to \citealt{1995ApJ...454..151M}). According to \citet{2008hsf1.book...36R}, the most massive star in Cyg OB3 is the O4I star HD 190429A with an age of  $\unit[1{.}6\pm1{.}0]{Myr}$ \citep[models from][]{1992A&AS...96..269S,2004A&A...424..919C} suggesting a younger age for Cyg OB3 or an age spread within the association. We therefore estimate the progenitor mass also for younger ages. However, from \autoref{tab:massage_Guitar} we can exclude very young ages of the pulsar's birthsite since the masses derived are larger than the upper limit for neutron star formation ($\approx\unit[30]{M_\odot}$, \citealt{2003ApJ...591..288H}), hence a black hole would have formed. Nonetheless, since, as mentioned above, Cyg OB3 consists of a number of smaller clusters with different ages ranging from $2$ to $\unit[\approx30]{Myr}$ (see Appendix of Paper I), it is well possible that the Guitar pulsar was born within an intermediate aged ($>\unit[8]{Myr}$) small cluster in Cyg OB3. Fitting isochrones from \citet{2002A&A...391..195G} to the HR diagram of members of Cyg OB3 listed by \citet{1978ApJS...38..309H}, we find an age range, namley that stars in Cyg OB3 formed from $6$ to $\unit[11]{Myr}$ ago. 

\begin{table}
\centering
\caption{Progenitor masses calculated for the difference between different ages for Cyg OB3 ($\mathrm{age_{\mathrm{CygOB3}}}$) and a pulsar age of $\unit[0{.}8]{Myr}$. Evolutionary models are taken from \citet{1980FCPh....5..287T} (T80), \citet{1989A&A...210..155M} (MM89) and \citet{1997PhDT........31K} (K97).}\label{tab:massage_Guitar}
\begin{tabular}{c c c c}
\toprule
$\mathrm{age_{\mathrm{CygOB3}}}$	&		T80		& MM89 	 & K97			\\
\multicolumn{1}{c}{[Myr]}					&		[$\mathrm{M_\odot}$] & [$\mathrm{M_\odot}$] & [$\mathrm{M_\odot}$]\\\midrule
$2$																&		$>50$	&	$>100$ & $>100$		\\
$5$																&		$46$	&	$47$	 & $42$			\\
$8$																&   $37$	&	$25$	 & $21$			\\
$12$															&		$24$	&	$16$	 & $15$			\\
\bottomrule
\end{tabular}
\end{table}


\section{Conclusions}

We have investigated the past trajectory of the Guitar pulsar (PSR B2224+65) as well as those of 140 OB associations and clusters (Paper I) to identify the parent association of the pulsar. Utilising statistical techniques we have taken the errors on the observables and, more importantly, a wide spectrum of radial velocities into account. However, the radial velocity may in principle be restricted to nearly zero due to the determination of the inclination of its bow shock \citep{2004ApJ...600L..51C}. Although we initially did not include constraints on the radial velocity, we find an association which fulfills this requirement. Hence, we conclude that the Guitar pulsar was born in the Cyg OB3 association $\approx\unit[0{.}8]{Myr}$ ago with its current radial velocity being $\approx\unit[-30]{km/s}$. Its kinematic age is slightly smaller than the characteristic age of $\unit[1{.}1]{Myr}$ derived from timing \citep{2004MNRAS.353.1311H} which is an upper limit for the pulsar age. For an age of Cyg OB3 of $\unit[8]{Myr}$ \citep{2001A&A...371..675U} we estimate a progenitor mass of $21$ to $\unit[37]{M_{\odot}}$ (O6 to O9.5 on the main sequence, \citealt{Schmidt-Kaler1982}). 

\section*{Acknowledgments}

We are grateful to Gracjan Maciejewski who re-investigated the age of Cyg OB3 for us. Furthermore, we would like to thank Valeri V. Hambaryan who guided our attention towards the Guitar pulsar. MMH and NT acknowledge
partial support from DFG in the SFB/TR-7 Gravitational
Wave Astronomy, RN acknowledges general support from the
German National Science Foundation (Deutsche Forschungsgemeinschaft,
DFG) in grants NE 515/23-1 and SFB/TR-7. The work of MMH and NT has been supported by CompStar, a research networking
 programme of the European Science Foundation (ESF). Our work has made use of the ATNF Pulsar Catalogue, version of February 2009 \citep{2005AJ....129.1993M}. 

\bibliographystyle{mn2e}
\bibliography{bib_GuitarPaper}

\label{lastpage}

\end{document}